\documentclass[%
twocolumn, prb, aps, superscriptaddress, longbibliography,showpacs,amsmath,amssymb,floatfix
]{revtex4-2}

\usepackage{changes}
\usepackage{graphicx}
\usepackage{dcolumn}
\usepackage{bm}
\usepackage{graphicx}    
\usepackage{amssymb}

\usepackage{amsmath}
\usepackage{epsfig}
\usepackage{xcolor}
\usepackage{tabu}
\usepackage{mathtools}
\usepackage[colorlinks,linkcolor=blue,anchorcolor=blue,citecolor=blue,urlcolor=blue]{hyperref}
\usepackage{physics}
\usepackage{float}
\usepackage{diagbox}
\usepackage{inputenc}

\begin{document}

\title{\textcolor{black}{Divergent density of states and non-analytic Lyapunov exponent in one-dimensional slowly varying systems}}
\author{Hai-Tao Hu}
\affiliation{Department of Physics, University of Science and Technology of China, Hefei, Anhui 230026, China}
\affiliation{Hefei National Laboratory, University of Science and Technology of China, Hefei 230088, China}
\affiliation{Eastern Institute of Technology, Ningbo 315200, China}
\author{Ming Gong}
\email{gongm@ustc.edu.cn}
\affiliation{Hefei National Laboratory, University of Science and Technology of China, Hefei 230088, China}
\affiliation{Key Laboratory of Quantum Information, University of Science and Technology of China, Hefei 230026, China}
\affiliation{Anhui Province Key Laboratory of Quantum Network, University of Science and Technology of China, Hefei 230026, China}
\affiliation{Synergetic Innovation Center of Quantum Information and Quantum Physics, University of Science and Technology of China, Hefei 230026, China}
\author{Guangcan Guo}
\affiliation{Hefei National Laboratory, University of Science and Technology of China, Hefei 230088, China}
\affiliation{Key Laboratory of Quantum Information, University of Science and Technology of China, Hefei 230026, China}
\affiliation{Anhui Province Key Laboratory of Quantum Network, University of Science and Technology of China, Hefei 230026, China}
\affiliation{Synergetic Innovation Center of Quantum Information and Quantum Physics, University of Science and Technology of China, Hefei 230026, China}
\author{Zijing Lin}
\email{zjlin@ustc.edu.cn}
\affiliation{Department of Physics, University of Science and Technology of China, Hefei, Anhui 230026, China}
\affiliation{Hefei National Laboratory, University of Science and Technology of China, Hefei 230088, China}
\date{\today}

\begin{abstract}
Localization of wave functions in disordered systems can be characterized by the Lyapunov exponent, which is zero in the extended phase and nonzero in the localized phase. Previous studies have shown that this exponent is an analytic function of eigenenergy in a given phase, thus its non-analytic behavior has been commonly used to determine the boundaries between the extended and localized phases. In this work, we show that if the localization centers are inhomogeneous across the whole chain and the system possesses (at least) two different localization modes, the Lyapunov exponent can become non-analytic in the localized phase at the boundaries between the different localization modes. We establish this central result by using several one-dimensional slowly varying models, and reveal that the non-analytic feature in the Lyapunov exponent is inherently tied to the singularities in the density of states through the Thouless formula. The possible existence of delicate structures in the localized phase effectively broadens our understanding of Anderson localization. 
\end{abstract}

\maketitle

\section{Introduction}\label{sec-introduction}
Anderson localization is one of the most fundamental concepts in condensed matter physics, which describes the suppression of wave transport caused by disorder-induced interference \cite{Anderson1958Diffusion, Thouless1974Electrons, Evers2008Anderson, 
Deng2019Quasicrystals, Loughi2019Topological, Roy2021Reentrant, Goncalves2023Critical, Longhi2024Dephasing, Chen2024Anderson}. Over the past decades, rapid experimental progress has enabled the controlled realization of disordered and quasiperiodic systems, allowing the localization–delocalization transition to be observed. These transitions have been experimentally observed in a broad range of experimental platforms, including cold atoms \cite{Semeghini2015Measurement, An2018Engineering,
Donald2020Observation, An2021Interactions, Zeng2024Transition}, optical systems \cite{Luschen2018Mobility, Goblot2020, Alexey2023Anderson}, and superconducting qubits \cite{Ghosh2014Simulating, Gong2021Experimental}. These experimental realizations have greatly enriched our understanding of how disorder shapes localization phenomena.

A central theoretical tool for describing Anderson localization is the Lyapunov exponent $\gamma(E)$, which characterizes the exponential decay of wave functions and thereby distinguishes extended states ($\gamma=0$) from localized ones ($\gamma>0$) \cite{Deych1998Statistics, Keselman2021Scrambling}. The energy dependence of $\gamma(E)$ encodes key information about the system's localization properties, serving as a powerful indicator of mobility edges (MEs) and phase boundaries. Several complementary methods have been developed to obtain $\gamma(E)$ and the associated MEs, such as the transfer-matrix approach \cite{Li2020Mobility, Lin2023critical, hu2025exact}, dual transformation method \cite{Biddle2010Mobility,Ganeshan2015Mobility, Liu2020Generalized, Wang2021Duality, hu2024hidden} and Avila's global theory \cite{Avila2015Global, Wang2020Quasiperiodic, Zhou2023Mobility}. In these studies, the Lyapunov exponent is assumed to be an analytic function of energy $E$ within the same phase, and the non-analytic transition in $\gamma(E)$ from zero to non-zero has been regarded as smoking-gun evidence of phase transition \cite{Zhang2022Lyapunov,
Garcia2024Sparsity, Ticea2024Stability, Li2024Asymmetric, wang2025exact}. However, with this wisdom, a fundamental question remains unexplored: can the Lyapunov exponent exhibit non-analytic behavior within a localized phase? At first glance, such behavior appears to contradict the standard picture, where non-analyticity of $\gamma(E)$ signals a boundary between distinct phases \cite{Wang2020Quasiperiodic}. Yet, there is no fundamental principle to forbid the non-analytic Lyapunov exponent in the localized phase, with $\gamma(E) > 0$. This suggests the possibility of much more subtle structures inside the localized phase than previously appreciated, which will greatly enhance our comprehension of Anderson localization in disordered models. 

In this manuscript, we explore this intriguing question by analyzing a class of one-dimensional slowly varying models that host multiple localization modes. We demonstrate that (I) the Lyapunov exponent $\gamma(E)$ can exhibit non-analytic behavior within the localized phase, without indicating a transition to an extended phase, and (II) this occurs when the system possesses multiple localization modes. We first show that a slowly varying potential which supports the localization around either potential nodes or extrema  naturally gives rise to two localization modes. We then prove the emergence of non-analytic behavior in $\gamma(E)$ at the boundaries separating these localization modes. A detailed analysis based on the Thouless formula 
\cite{Thouless1972relation} shows that the divergences in the density of states (DOS) play a crucial role in the non-analytic transition between localized states. Our results suggest that the localized phase may possess much more delicate structures than previously recognized, warranting more careful consideration in future studies and offering significant insight into our understanding of localization phenomena.

\textcolor{black}{The rest of this manuscript is organized as follows. Section~\ref{sec-potentials} takes slowly varying potentials as a representative example to illustrate that the non-analytic Lyapunov exponent arises from the coexistence of two kinds of localized modes in a localized phase. Section~\ref{sec-general} extends this analysis to more general slowly varying models and presents additional supporting evidence. Finally, Sec.~\ref{sec-conclusions} summarizes and discusses our results.}

\begin{figure}
\includegraphics[width=0.48\textwidth]{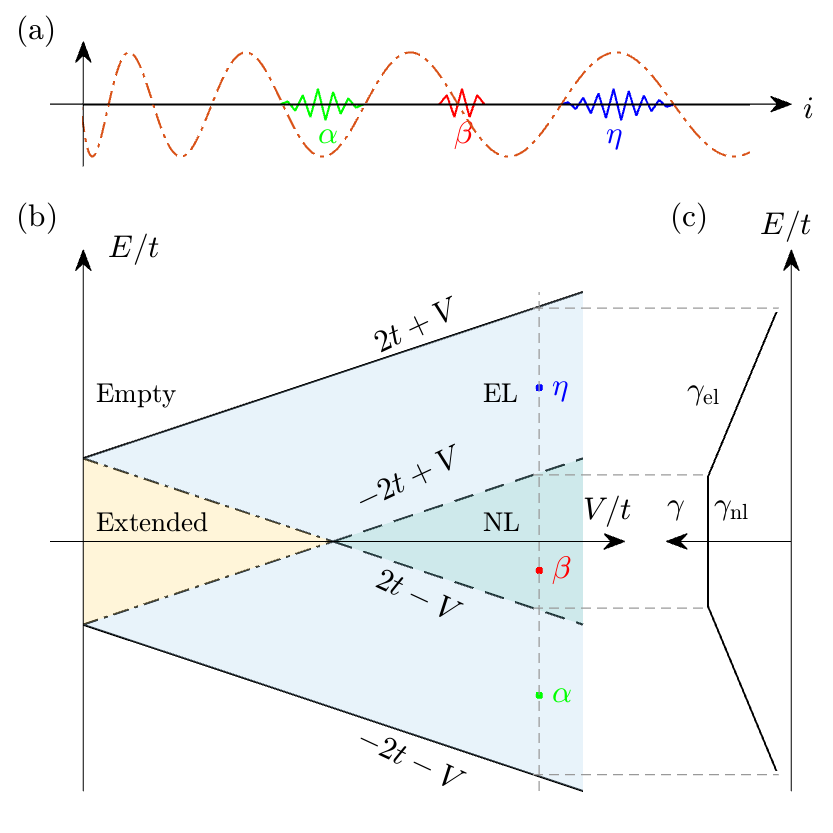}
\caption{(a) Two distinct localization modes in the slowly varying potential [see Eq.~(\ref{eq-hami-dia})]. Symbols represent wave function localized at the nodes and extrema of the potential. (b) Phase diagram of the model. The labels 'Extended', 'NL', 'EL', and 'Empty' indicate extended states, node localized states, extremum localized states, and absence of states, respectively. (c) Schematic Lyapunov exponents for NL and EL, denoted as $\gamma_\text{nl}$ and $\gamma_\text{{el}}$ respectively.}
\label{fig-fig1}
\end{figure}

\section{Slowly varying potential model}
\label{sec-potentials}

\subsection{Dual localization modes}
We begin by analyzing the localization properties of conventional disordered systems, which is helpful for understanding the non-analytic behavior of the Lyapunov exponent. In these models, the on-site potentials can be regarded as homogeneous in the sense of ergodicity, thus the localization centers of the wave functions will be uniformly distributed in the whole chain. Consequently, these models exhibit only a single localization mode, implying that the Lyapunov exponent $\gamma(E)$ should be an analytic function of $E$ \cite{Wang2020Quasiperiodic}. This idea further suggests that for $\gamma(E)$ to be a non-analytic function of $E$ in the localized phase, the system must possess at least two distinct localization modes, such as different localization centers. We implement this concept through the following slowly varying potential model \cite{Thouless1988Localization, Das1988Mobility, Griniasty1988Localization, Das1990Localization, Liu2017Fate, Lu2022Exact,Tang2024Mobility} 
\begin{align}
\mathcal{H} = -t \sum_{i=1}^{L-1} (c_{i + 1}^\dagger c_i + \mathrm{H.c.} ) + \sum_{i=1}^L V_i c_i^\dagger c_i,
\label{eq-hami-dia}
\end{align}
where $c_i^\dagger$ ($c_i$) denotes the creation (annihilation) operator at site $i$, $t$ represents the nearest-neighbor hopping term, and $L$ is the chain length. If the potential is taken as $V_i =V\cos(\pi \alpha i^\nu)$ with $ 0 < \nu < 1$ \cite{Thouless1988Localization, Das1988Mobility, Griniasty1988Localization,Das1990Localization, Liu2017Fate, Lu2022Exact, Tang2024Mobility}, there exist two distinct localization modes as shown in Fig.~\ref{fig-fig1}(a), that is, the wave functions are likely to be localized either at potential nodes or potential extrema when $V$ is sufficiently large. In this way, the localization centers (mean position of wave function) exhibit spatial inhomogeneity, marking the emergence of dual localization centers in the system. The inhomogeneity of the localization centers is the reason for the non-analytic behavior of Lyapunov exponent.

The form of $V_i$ allows for the constant approximations in the large-$i$ limit, facilitating analytic treatment of localization properties. In the bare hopping regime ($V = 0$), the kinetic energy is bounded by $|K| \leq 2t$ (with $t > 0$). Consequently, when the wave functions are localized at the potential extrema, their energy should be either $K - V$ or $K + V$, defining the boundary curves $\pm 2t \pm V$, which intersect at $V = 2t$. In contrast, when states are fully localized at the potential nodes, their energy is expected to be independent of the potential depth, leading to $E \approx 0$. It will depend on $V$ when the wave functions are spatially localized near the nodes. This argument yields the phase diagram shown in Fig.~\ref{fig-fig1}(b). While the boundaries between the extended and localized states seen in Fig.~\ref{fig-fig1}(b) have been rigorously established in Ref. \cite{ Das1990Localization}, the key insight here is that the presence of two distinct localization modes allows the localization length and the associated Lyapunov exponent to exhibit non-analytic variations as functions of energy $E$ in the localized phase. More specifically, if we denote the Lyapunov exponents in the two localization modes as $\gamma_\text{nl}(E)$ and $\gamma_\text{el}(E)$, the non-analytic behavior is bound to occur at $\gamma_\text{nl}(E) = \gamma_\text{el}(E)$, as illustrated in Fig.~\ref{fig-fig1}(c). It should be emphasized that these states belong to the same phase in the sense that their measured properties, such as conductance will be unchanged in the thermodynamic limit. From a broader perspective, this phenomenon highlights that even within a single phase characterized by specific order parameters, certain physical properties may still exhibit some non-analytic behaviors.  

\subsection{Lyapunov exponent, DOS, and Thouless formula}
According to the Thouless formula, the Lyapunov exponent can be obtained by \cite{Thouless1972relation, Acharya2024Localization}
\begin{align}
\gamma(E) = \int dE' D(E') \ln |E-E'|,
\end{align}
where the DOS is defined as 
\begin{align}
D(E) = \sum_{n = 1}^L \delta(E-E_n).
\end{align}
This relation, termed as Hilbert transformation, implies that singularities in $D(E)$ can yield non-analytic behavior in $\gamma(E)$. This behavior can be categorized into two distinct cases, one with $\gamma(E)$ changing from zero to a finite value corresponding to Anderson transition \cite{Das1988Mobility}, and the other with $\gamma(E)$ varying between two finite values representing transition between different localization modes. For the Aubry-André-Harper (AAH) model \cite{Aubry1980Analyticity, Kraus2012Topological, cai2013topological, ganeshan2013topological}, $\gamma(E)$ remains zero for all energies $E$ when $V<2t$, while for $|V| > 2t$, it is given by $\gamma = \text{ln}(|V/2t|)$, indicating a phase transition at $V = \pm 2t$, as determined by self-duality \cite{Aubry1980Analyticity}. Consequently, although $\gamma(E)$ remains continuous as a function of $E$, its first derivative $\partial \gamma$ exhibits discontinuities at the phase boundaries, signaling a phase transition. This constitutes a major result of this work, that is, the Lyapunov exponent can be a continuous yet non-analytic function of $E$ in the localized phase, which always indicates a transition between different kinds of states.

\textcolor{black}{
\subsection{Methods for characterizing localization properties}
Here, we summarize the numerical methods used to characterize the localization properties of the system, including the fractal dimension and the Lyapunov exponent.
\subsubsection{Fractal dimension $D_2$}
In the theory of Anderson localization, the inverse participation ratio (IPR) is a standard measure to quantify the localization properties of eigenstates, defined as 
\begin{align}
{\rm IPR}^{n} = \sum_{i = 1}^{N} |\psi_{i}^{n}|^4,
\end{align}
Here, $\psi_{i}^{n}$ is the normalized wavefunction at site $i$, with $n$ being the band index. In the limiting cases, ${\rm IPR}^{n}=1$ for a state localized on a single site ($L^{0}$ scaling), while ${\rm IPR}^{n}\propto L^{-1}$ for a uniformly extended state. Based on this, we can define the fractal dimension as
\begin{align}
D_2^{n} = - \lim_{L \rightarrow \infty} \frac{{\rm ln}{\rm IPR}^{n}}{{\rm ln}L}
,
\end{align}
which falls between $0$ and $1$. Here, $D_2$ tends to 1 for extended states and 0 for localized states, while $0 < D_2 < 1$ indicates multifractal states. 
\subsubsection{Lyapunov exponent $\gamma$}
The Lyapunov exponent of a tight-binding model involving only nearest-neighbor hopping terms can be calculated as 
\begin{align}
\gamma(E) = \lim \limits_{L \rightarrow \infty} \frac{1}{2L} \ln (\mathbf{T}_L^\dagger \mathbf{T}_L), \hspace{5mm} \mathbf{T}_L = \prod \limits_{i = 1}^N T_i,
 \label{eq11}
\end{align}
where the transfer matrix $T_i$ is
\begin{align}
T_i =
\begin{pmatrix}
{V_i - E} & -1 \\
1 & 0
\end{pmatrix}
. \label{eq10}
\end{align}
Here, $\mathbf{T}_L$ is the product of single-site transfer matrices $T_i$ up to length $L$. In practice, the QR decomposition is employed to improve the numerical stability and efficiency in computing $\mathbf{T}_L$.}

\begin{figure}
\includegraphics[width=0.48\textwidth]{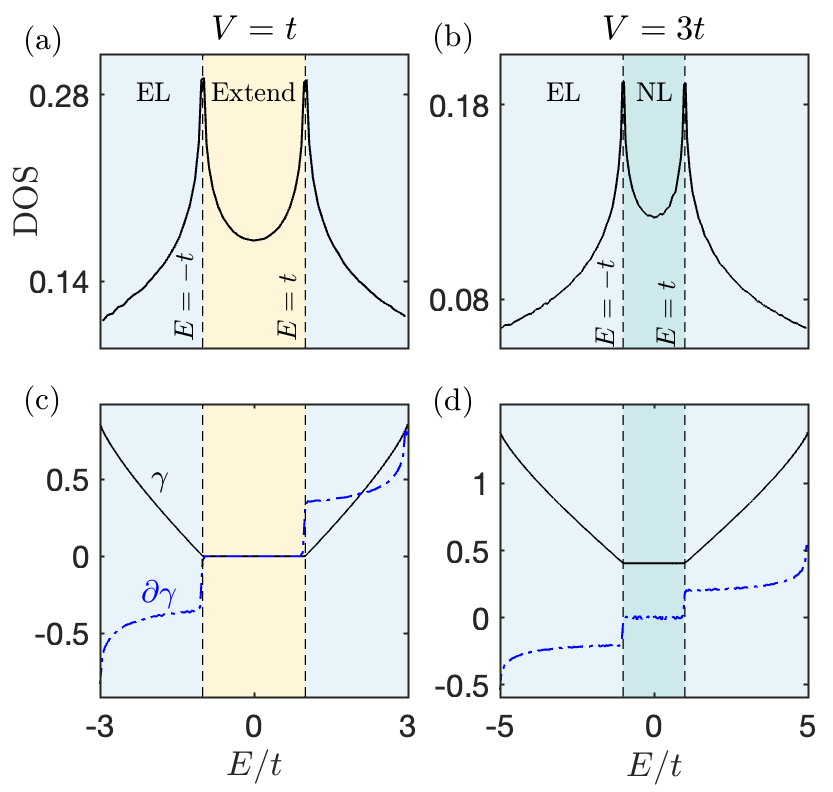}
\caption{(a)-(b) DOS, and (c)-(d) Lyapunov exponent $\gamma$ and its derivative $\partial \gamma$ as functions of energy $E$ for various potential strength $V$. The potential is given by $V_i = V\cos(\pi \alpha i^\nu)$. The dashed lines denote the boundaries $E = \pm(2t-V)$, which separate the extended states from the localized states when $V<2t$, and distinguish between two kinds of localized states when $V>2t$.}
\label{fig-fig2}
\end{figure}

\begin{figure}
\includegraphics[width=0.48\textwidth]{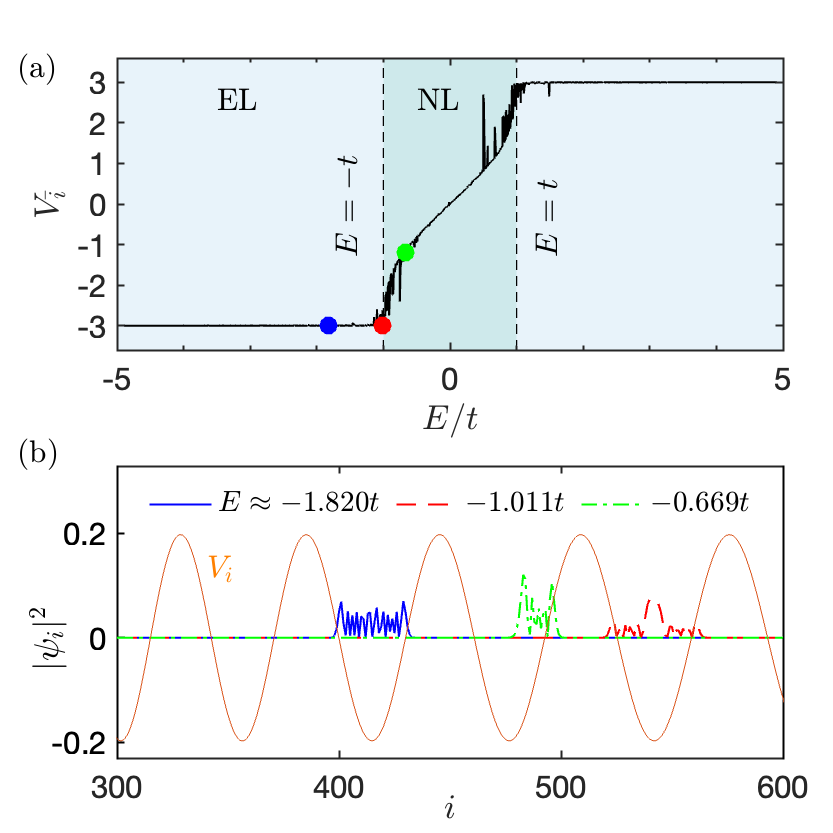}
\caption{(a) The potential at the localization center $f_\nu(\bar{i})$ versus energy $E$ when $V =3t$. The dashed lines denote the boundaries between two localized modes at $E = \pm t$. (b) Spatial distribution of three representative
eigenstates [as marked in Fig.~\ref{fig-fig3}(a)]. The orange solid line represents the potential.}
\label{fig-fig3}
\end{figure}

\subsection{Numerical results}
For the slowly varying potential $V_i = V\cos(\pi \alpha i^\nu)$, the numerical results for the DOS, $\gamma(E)$ and its first derivative $\partial \gamma$ at $V = t$ and $V = 3t$ are presented in Fig.~\ref{fig-fig2}. When $V = t$, the DOS diverges at $E = \pm t$ [see Fig.~\ref{fig-fig2}(a)], with the Lyapunov exponent remaining zero for $|E| < t$, characterizing an extended phase [see Fig.~\ref{fig-fig2}(b)]. For $|E|>t$, the Lyapunov exponent approximately follows $\gamma \propto |E|-t$ around the critical points $E_{\rm c} = \pm t$. These results indicate that the phase transition, characterized by non-analytic behavior of $\gamma$, is driven by the divergence of DOS. When $V = 3t$, all eigenstates are localized with $\gamma > 0$ [see the black solid line in Fig.~\ref{fig-fig2}(d)], and the DOS still diverges at $E = \pm t$ [see the black solid line in Fig.~\ref{fig-fig2}(b)]. Meanwhile, we still observe non-analyticity of $\partial\gamma$ in this case. All of the evidence indicates the existence of two different localized states.  

To gain deeper insight into the nature of different localization modes, we calculate the localization center as $\bar{i} = \sum_i i |\psi_i|^2$, where $\psi$ denotes a normalized eigenstate. For node localized states, we expect $V_{\bar{i}} \sim 0$, whereas for extreme localized states, we expect $V_{\bar{i}} \sim \pm V$. Focusing on the regime $V>2t$, we set $V = 3t$ and plot the potential at the localization center as a function of $E$ in Fig.~\ref{fig-fig3}(a). In this case, the boundaries between these localized states occur at $E = \pm t$, which agree well with our expectations. For extreme localized states, we find $V_{\bar{i}} \rightarrow \pm V$, while for node localized states, we observe $|V_{\bar{i}}| < V$. Three typical wave functions corresponding to these localization modes are presented in Fig. \ref{fig-fig3}(b). These results reveal that the boundaries between two localization modes give rise to the non-analyticity of $\gamma(E)$, providing a physical interpretation of this phenomenon from the perspective of wave function. 

\begin{figure}
\includegraphics[width=0.48\textwidth]{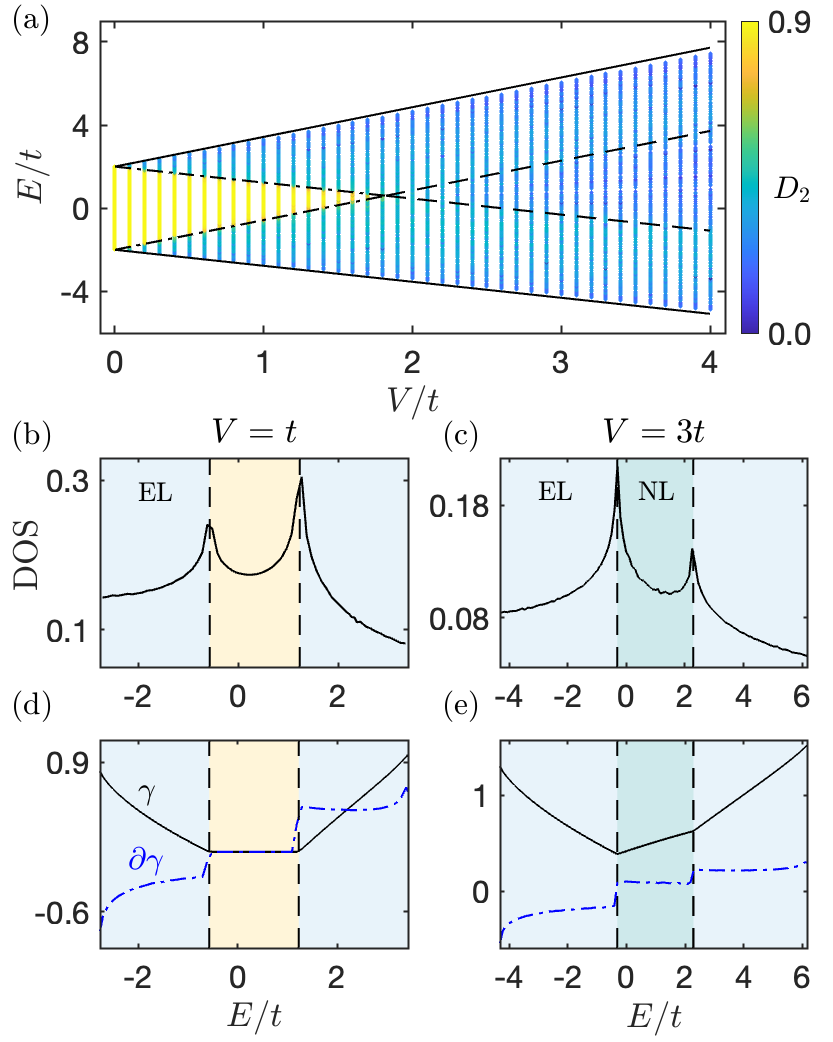}
\caption{\textcolor{black}{(a) Fractal dimension $D_2$ versus eigenenergy $E$ and hopping strength $\lambda$ for the slowly varying model  described by Eq.~(\ref{eq-model2}), with $b = 0.3$. (b)-(c) DOS, and (d)-(e) Lyapunov exponent $\gamma$ and its derivative $\partial \gamma$ versus $E$ for different $V$. The dashed lines in (b) and (d) represent the boundaries $E = -4t/7$ and $E =16t/13$, whereas in (d) and (f), the boundaries are located at $E = -4t/13$ and $E =16t/7$.}}
\label{fig-fig6}
\end{figure}

The above results may be found in much more general slowly varying potentials. To this end, we have also verified that when the on-site potential is given by  \cite{Ganeshan2015Mobility, Liu2022Anomalous, Padhan2024Complete, Duncan2024Critical}
\begin{equation}
V_i = V {\cos(\pi \alpha i^{\nu}) \over 1 - b \cos(\pi \alpha i^{\nu})},
\label{eq-model2}
\end{equation}
with $|b| < 1$ and $0 < \nu < 1$, the same non-analyticity of $\gamma(E)$ can be obtained. Using the same approach, we find that the phase boundaries can be described by $E = \pm 2t + V/(1-b)$ and $E = \pm 2t - V/(1+b)$. In this way, all states are localized when $V > V_c = 2t(1-b^2)$, and the boundaries between node and extreme localized states occur at $E = - 2t + V/(1 - b)$ and $E = 2t - V/(1+b)$ when $V > V_c$. The singularities in DOS lead to the non-analytic behavior of the Lyapunov exponents in the localized phase. \textcolor{black}{The corresponding numerical results is presented in Fig.~\ref{fig-fig6}, which further demonstrate that the emergence of non-analytic $\gamma(E)$ is a general feature of slowly varying potentials supporting multiple localization modes.}

\section{Slowly varying hopping model}\label{sec-general}
To further generalize our analysis, we consider a model with slowly varying hopping, described by the Hamiltonian \cite{Liu2017Fate}
\begin{align}
\mathcal{H} = - \sum_{i=1}^{L-1} g_\nu(i) (c_{i + 1}^\dagger c_i + \mathrm{H.c.} ) + \sum_{i=1}^L V c_i^\dagger c_i,
\label{eq-hami-offdia}
\end{align}
where $g_\nu(i)$ denotes the quasiperiodic slowly varying hopping term and $V$ represents the on-site potential. We define $g_\nu(i)$ as
\begin{align}
g_\nu(i) = t + \lambda \frac{\cos(\pi \alpha i^\nu)}{1-b\cos(\pi \alpha i^\nu)},
\label{eq-gnu}
\end{align}
where $t$ is a constant, and $\lambda$ characterizes the amplitude of the slowly varying hopping modulation. When $|\lambda| < (1+b)t$, all sites are connected, allowing the existence of extended states. However, when $|\lambda| > (1+b)t$, there exist infinitely many indices $i$ satisfying $|g_\nu(i)| < \varepsilon$, where $\varepsilon$ is an infinitesimally small positive constant. As a result, the hopping amplitudes vanish locally, dividing the chain into weakly coupled blocks and giving rise to spatially localized states. Similar to the slowly varying potential model, this system supports two distinct localization modes: node localization at $g_\nu(i) \sim 0$ and extreme localization at $g_\nu(i) \sim t + \lambda/(1-b)$. Following the similar approach \cite{Das1990Localization, Liu2017Fate, Tang2024Mobility}, we derive the boundaries separating these localization modes as 
\begin{align}
E_c = \pm2(t-\frac{\lambda}{1+b}) + V.
\label{eq-boundaries}
\end{align}

\begin{figure}
\includegraphics[width=0.48\textwidth]{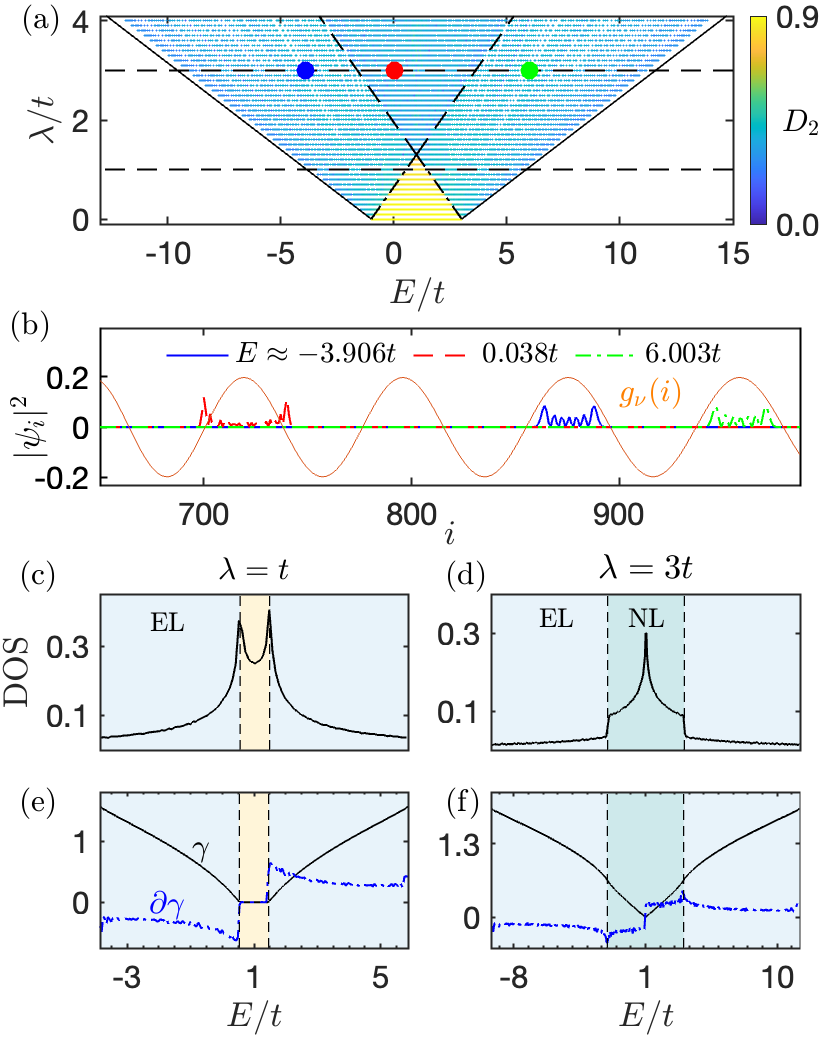}
\caption{(a) Fractal dimension $D_2$ versus eigenenergy $E$ and hopping strength $\lambda$ for the quasiperiodic slowly varying hopping model  described by Eq.~(\ref{eq-gnu}), with $b = 0.3$. (b) Three representative eigenstates as marked in Fig.~\ref{fig-fig4}(a). (c)-(d) DOS, and (e)-(f) Lyapunov exponent $\gamma$ and its derivative $\partial \gamma$ versus $E$ for different $\lambda$. The dashed lines in (c) and (e) represent the boundaries $E = 7t/13$ and $E =19t/13$, whereas in (d) and (f), the boundaries are located at $E = -21t/13$ and $E =47t/13$.}
\label{fig-fig4}
\end{figure}

The phase diagram obtained from numerical simulations is presented in Fig.~\ref{fig-fig4}(a). In Fig.~\ref{fig-fig4}(b), we demonstrate that these states are indeed spatially localized at distinct localization centers. Meanwhile, we calculate the DOS for $V = t$ and $V = 3t$, as shown in Figs.~\ref{fig-fig4}(c) and \ref{fig-fig4}(d). From the DOS, we can determine the Lyapunov exponent $\gamma(E)$ and its first derivative $\partial\gamma$, which are presented in Figs.~\ref{fig-fig4}(e) and \ref{fig-fig4}(f). For $V = t$, we find that $\gamma(E) = 0$ when $7t/13< E < 19t/13$, indicating a localized-delocalized transition driven by the divergence of DOS. When $V = 3t$, $\gamma(E)$ remains positive, indicating the localization of all states. We find discontinuous DOS and non-analytic Lyapunov exponents at $E = -21t/13$ and $E = 47t/13$.

\begin{figure}
\includegraphics[width=0.48\textwidth]{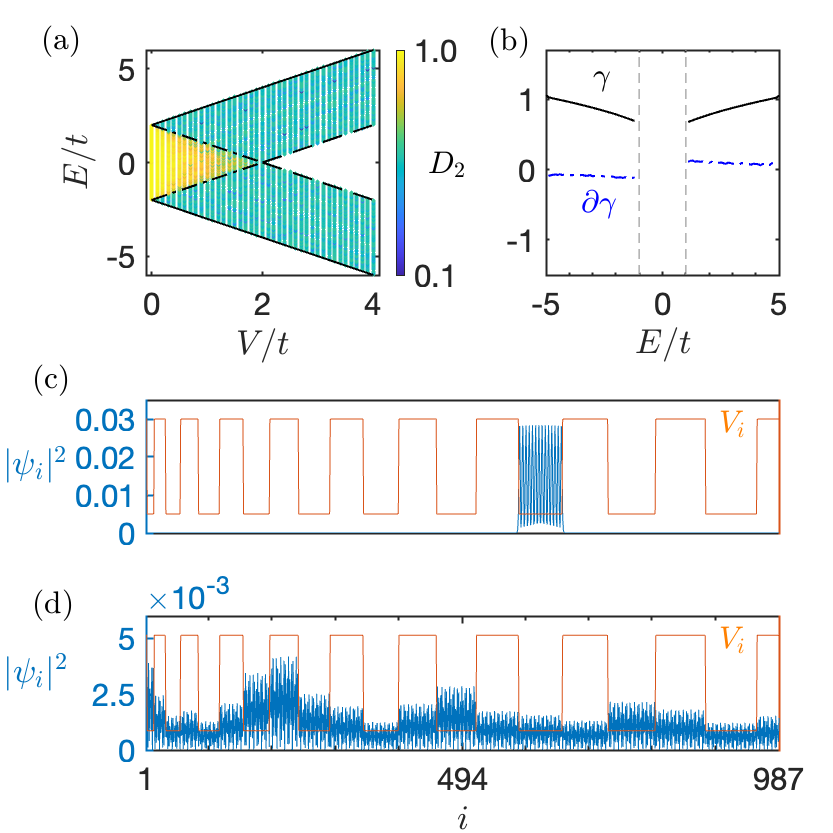}
\caption{(a) Fractal dimension $D_2$ as a function of eigenenergy $E$ and potential strength $V$ for the quasiperiodic slowly varying square-well potential model. (b) Lyapunov exponent $\gamma$ and its derivative $\partial \gamma$ versus $E$ when $V = 3t$. The dashed lines denote the boundaries $E=\pm t$. (c) A representative localized state with $E \approx -2.12t$, and (d) a representative extended state with $E \approx 0$ when $V = 0.5t$. The orange solid line denotes the potential.}
\label{fig-fig5}
\end{figure}

Building on the above examples and insights, we arrive at the central conclusion of this study. In the following, we provide a counterexample to further substantiate this conclusion. Specifically, we consider a model characterized by the following slowly varying square-well potential \cite{Das1988Mobility, Das1990Localization}
\begin{align}
V_i &= V \cdot \text{sign}[\cos(\pi \alpha i^\nu)],
\end{align}
where $\text{sign}(x)$ denotes the sign function. Since this potential lacks nodes, it does not support node localization, yielding a single localization mode. To verify this, we plot the fractal dimension $D_2$ versus energy $E$ and potential strength $V$ in Fig.~\ref{fig-fig5}(a), with phase boundaries determined by $E = \pm(2t \pm V)$. We confirm that in the overlapped regime for these four curves, the wave functions are extended, whereas in the non-overlapped regime, wave functions are localized at the potential extrema as illustrated in Figs.~\ref{fig-fig5}(c) and \ref{fig-fig5}(d). Additionally, we calculate the DOS and the corresponding Lyapunov exponent, demonstrating that both $\gamma(E)$ and its first-order derivative are analytic functions of $E$, as presented in Fig.~\ref{fig-fig5}(b). Since this model only supports a single localization mode, the non-analytic behavior in $\gamma(E)$ is absent, leading to an analytic Lyapunov exponent similar to those observed in disordered and quasiperiodic potential models \cite{Li2020Mobility}. 

\section{Conclusions}\label{sec-conclusions}
The localization of wave functions in disordered potentials can be characterized by the Lyapunov exponent $\gamma(E)$, whose transition from zero to a finite value marks the change from extended to localized states \cite{Wang2024Non, Cai2025Origin,zhou2025fundamental}. Although $\gamma(E)$ varies continuously with energy, it is generally non-analytic across localization transitions. In this work, we address the fundamental question of whether $\gamma(E)$ can be non-analytic in the localized phase, which is not prohibited by its definition. We propose an intuitive mechanism for this phenomenon that when the system supports two (or more) kinds of localized states, the Lyapunov exponent can display non-analytic behavior within the localized phase. To illustrate this idea, we analyzed one-dimensional models with slowly varying potential \cite{Thouless1988Localization, Das1988Mobility, Griniasty1988Localization, Das1990Localization, Liu2017Fate}, where we observe a non-analytic Lyapunov exponent in the localized phase. Further investigation shows that these systems support two kinds of localized states which localize at the potential extrema and potential nodes, and the boundaries for these localized states are the transition points of the non-analytic Lyapunov exponent. According to the Thouless formula, the non-analyticity of the Lyapunov exponent is attributed to the singularities of the DOS \cite{Thouless1972relation}. Finally, we extend our discussion to one-dimensional models with slowly varying hopping and find similar phenomena. In this work we tentatively consider all localized states to be the same phase, and in future, it is quite possible that if new classification criteria are invented, these states may also be categorized into different phases. Since this non-analyticity of the Lyapunov exponent is in principle allowed in theory as demonstrated in this manuscript, this idea may be generalized to higher dimensions, which should greatly broaden our understanding of Anderson localization \cite{Anderson1958Diffusion, Thouless1974Electrons, Evers2008Anderson}. 

\begin{acknowledgements}
We thank Profs. Zhenyu Zhang, Qian Niu, and Qi Zhou for valuable discussion. This work is supported by the National Natural Science Foundation of China (Grant No. 12374017, No. 12074362 and  No. U23A2074), the Innovation Program for Quantum Science and Technology (2021ZD0303303, 2021ZD0301200, 2021ZD0301500) and the Strategic Priority Research Program of the Chinese Academy of Sciences (Grant No. XDB0500000).
\end{acknowledgements}

\bibliography{ref.bib}

\end{document}